\documentclass[iop,apj]{emulateapj}
\usepackage{natbib}
\usepackage{color}
\usepackage{graphicx}
\usepackage{amsmath,amsfonts}
\newcommand\etal{{\em et al. }}

\newcommand\1{$^1$}
\newcommand\2{$^2$}
\newcommand\3{$^3$}
\newcommand\4{$^4$}
\newcommand\5{$^5$}

\begin{document}

\title{Excess Optical Enhancement Observed with ARCONS for Early Crab Giant Pulses} 

\shorttitle{Crab with ARCONS}

\author{M.J. Strader\1, M.D. Johnson\2, B.A. Mazin\1, G.V. Spiro Jaeger\1, C.R. Gwinn\1, S.R. Meeker\1, P. Szypryt\1, J.C. van Eyken\1,  D. Marsden\1, K. O'Brien\3, A.B. Walter\1,  G. Ulbricht\1, C. Stoughton\4, and B. Bumble\5}

\affil{\1 Department of Physics, University of California, Santa Barbara, CA 93106, USA}
\affil{\2 Harvard-Smithsonian Center for Astrophysics, 60 Garden Street, Cambridge, MA 02138, USA}
\affil{\3 Department of Physics, University of Oxford, Denys Wilkinson Building, Keble Road, Oxford, OX1 3RH, UK}
\affil{\4 Fermilab Center for Particle Astrophysics, Batavia, IL 60510, USA}
\affil{\5 NASA Jet Propulsion Laboratory, 4800 Oak Grove Drive, Pasadena, CA 91125, USA}
\shortauthors{Strader \etal}

%\vskip 1 truein

\begin{abstract}
We observe an extraordinary link in the Crab pulsar between the enhancement of an optical pulse and the timing of the corresponding giant radio pulse.  At optical through infrared wavelengths, our observations use the high time resolution of ARCONS, a unique superconducting energy-resolving photon-counting array at the Palomar 200-inch telescope.  At radio wavelengths, we observe with the Robert C. Byrd Green Bank Telescope and the GUPPI backend.  We see an $11.3\pm2.5\%$ increase in peak optical flux for pulses that have an accompanying giant radio pulse arriving near the peak of the optical main pulse, in contrast to a $3.2\pm0.5\%$ increase when an accompanying giant radio pulse arrives soon after the optical peak. We also observe that the peak of the optical main pulse is $2.8\pm0.8\%$ enhanced when there is a giant radio pulse accompanying the optical interpulse.  We observe no statistically significant spectral differences between optical pulses accompanied by and not accompanied by giant radio pulses. Our results extend previous observations of optical-radio correlation to the time and spectral domains.  Our refined temporal correlation suggests that optical and radio emission are indeed causally linked, and the lack of spectral differences suggests that the same mechanism is responsible for all optical emission.
\end{abstract}


\section{Introduction}

\subsection{The Crab Pulsar}

The Crab pulsar is among the youngest known pulsars.  Pulsed emission from the Crab has been observed at most spectral ranges detectable, from dekameter radio wavelengths to gamma-ray energies greater than 100 GeV \citep{ref:Mutel1974,ref:Kuzmin2002,ref:Veritas2011}.  A nearly-orthogonal rotator, it displays both a main pulse and a interpulse.  The close alignment in time of the Crab pulsar's emission from radio to gamma energies (after removing the appropriate dispersive delay) suggests that a single emission region gives rise to the entire spectrum. The detection of pulsed emission above 100 GeV by VERITAS implies an emission altitude of at least 100 km above the neutron star's surface \citep{ref:Veritas2011}. The optical lightcurve is broader in time than the radio, but shows nearly concurrent peaks corresponding to the main pulse and the interpulse.

Most remarkably, and unlike the majority of pulsars, its radio flux is dominated by a small number of giant pulses, which regularly have flux densities a thousand times the mean radio emission \citep{ref:Heiles1970,ref:Staelin1970}. The occurrence of these giant radio pulses (GRPs) is random, but they appear in a narrow range of pulse phase nearly coincident with the peaks of the main pulse and interpulse, although between 4-8.4 GHz, giant pulses also occur at two additional phases that are not near optical peaks. Giant pulses show variability on nanosecond timescales \citep{ref:Hankins2003} and appear to follow a power-law distribution in amplitude (with power law indices of roughly -2 to -3), with possible deviation at the highest amplitudes \citep{ref:Argyle1975,ref:Mickaliger2012}.  

\subsection{Multi-Frequency Observations}

Pulsar emission at optical and X-ray wavelengths is commonly ascribed to incoherent synchrotron radiation \citep{ref:Harding2008}.  Radio photons play an essential role in catalyzing emission of this radiation \citep{ref:Petrova2009}.  This predicts a link between radio emission and higher energy emission.  Accordingly, the few associations of emission in different spectral bands that can be observed have an important role in constraining models of pulsar emission.  For example, \citet{ref:Lommen2007} found that when the radio pulse of the Vela pulsar arrives early, there is an enhancement in the immediately following X-ray peak, although shot noise from the X-ray emission from the surrounding Vela-X supernova remnant led to increased uncertainty in the X-ray fluxes of individual pulses.  Repeated attempts to detect correlations between the radio and gamma emission of the Crab giant pulses have not uncovered any statistically significant correlation \citep{ref:Lundgren1995,ref:Bilous2011}.

An optical-radio correlation in the Crab pulsar has been detected, however.  Optical pulses have been seen to be 3\% brighter when accompanied by a GRP \citep{ref:Shearer2003,ref:Collins2012}.  Such enhancement, combined with the lack of radio-gamma correlation, could reflect rapid (${<}10\ \mu\mathrm{s}$) local density variations in the plasma stream, which would affect the coherent (radio) emission but not the incoherent (gamma) emission \citep{ref:Hankins2003}. 

We extend and refine measurements of this correlation with simultaneous observations of the Crab pulsar using the GUPPI backend \citep{ref:DuPlain2008_GUPPI} of the Robert C. Byrd Green Bank Telescope (GBT) and a new optical through near-infrared instrument, the ARray Camera for Optical to Near-IR Spectrophotometry (ARCONS) \citep{ref:Mazin2013}, at the 200-inch Hale telescope.

\section{Observations}

\subsection{Radio Data and Analysis}

We observed the Crab pulsar for two hours on 12 December 2012 using the recently built Green Bank Ultimate Pulsar Processing Instrument (GUPPI) at the GBT.\footnote{The National Radio Astronomy Observatory is a facility of the National Science Foundation operated under cooperative agreement by Associated Universities, Inc.} GUPPI was used in 800 MHz coherent search mode centered on 1.5 GHz. 

We developed a timing model for the observation using the software package TEMPO2 \citep{ref:Tempo2}. However, we adopted the dispersion measure (DM) reported in the Jodrell Bank Crab pulsar monthly ephemeris \citep{ref:Lyne1993}.\footnote{http://www.jb.man.ac.uk/pulsar/crab.html} This DM has an associated uncertainty of $0.005\mathrm{\ pc/cm}^3$, equivalent to a radio timing uncertainty of ${\sim} 10\ \mu\mathrm{s}$, which is the dominant source of error in the radio side of our timing comparisons.

Next, we analyzed single-pulse properties using custom software that utilized libraries from PSRCHIVE \citep{ref:Hotan2004_Psrchive}. To eliminate pulses that were contaminated by sporadic, impulsive interference, we excluded any pulse from analysis that had a maximal band-averaged intensity in the off-pulse region exceeding the off-pulse mean by greater than five times the standard deviation of all the off-pulse samples. Approximately $0.3\%$ of pulses were eliminated on the basis of this criterion. For the remaining pulses, we tabulated the mean and maximum intensity of the main and interpulse phase ranges as well as the arrival phase of the respective points of maximum intensity. Subsequently, we used TEMPO2 to calculate the barycentric arrival time of each pulse.  

To establish an approximate flux calibration, we utilize the Crab Nebula. At 1.5 GHz, the Crab Nebula is both bright and unresolved by the GBT. We therefore estimate a contribution of $955 \nu_{\rm GHz}^{-0.27}\mathrm{\ Jy} \approx 850 \mathrm{\ Jy}$ \citep{ref:Bietenholz1997}, which easily dominates the system noise. We thus equate this contribution with the average of the off-pulse region to obtain an approximate conversion from our measured flux density to Jy.

Observed radio pulses that had a peak flux density above 28.4 Jy and that arrived during the phase range 0.9915 to 1.0042 (where the phase of the peak of the average radio main pulse has been defined as 1) were tagged as main pulse GRPs.  There were 7205 of these observed.  Radio pulses with a flux density above 25.1 Jy that arrived in the phase range 0.3928 to 0.4099 were tagged as interpulse GRPs.  In addition, interpulse GRPs that occur in the same period as a main pulse GRP were excluded, so that enhancement by main pulse GRPs does not affect the interpulse result. There were 2237 pulse detections meeting these requirements.  The flux cutoffs were chosen to exclude the significant number of false positives at lower fluxes, which would artificially lower the average optical enhancement calculated, as discussed in Section \ref{sec:radioflux}.  The phase constraints were chosen to include nearly all GRP detections while minimizing the number of false GRP detections.  Outside of the chosen phase ranges, the fraction of false positives becomes significant and, similar to low flux GRPs mentioned above, the enhancement computed at these phases would be artificially lowered. Even with the flux and phase restrictions roughly 3\% of the pulses tagged as main pulse GRPs and 60\% of the interpulses are expected to be false detections.

\subsection{ARCONS}
ARCONS is the first of a new generation of astronomical instrumentation that uses Microwave Kinetic Inductance Detectors (MKIDs), which enable time-tagging and energy-tagging of individual photons \citep{ref:Day2003,ref:Mazin2013}.  This capability enabled us to observe correlations in optical flux and spectrum with the occurrence of GRPs as well as correlations with other parameters, such as the arrival phase of GRPs.  

MKIDs enable the detection of individual photons with high time resolution (at the $\ \mu\mathrm{s}$ level) and with simultaneous energy resolution ($R=E/\Delta E \sim 8$ at 4000 \AA). MKIDs do not suffer from read noise or dark current.  This instrument provides vast improvements over traditional CCD detectors when observing transient events, such as giant pulses.  The MKID array used for these observations consisted of 2024 (46x44) pixels.  The instrument's plate scale is 0.45"/pixel, so the field of view is 20"x20".  It is sensitive to wavelengths in the range 4000-11000 \AA.

\subsection{Optical Data and Analysis}

The Crab pulsar was observed in the optical from the Coud\'{e} focus of the 200-inch Hale Telescope at Palomar Observatory from 03:30 to 06:37 AM on 12 December 2012 UTC.  The seeing fluctuated between 1" and 1.5" during the observation.

The optical data were compiled into photon lists by the ARCONS pipeline \citep{ref:Mazin2013}.  Each photon was tagged during the observation with a timestamp derived from a Stanford Research FS725 Rubidium 10 MHz frequency standard, which was synced with the 1 pulse per second (PPS) output of a Meinberg GPS170PEX GPS board.  The timestamps were later corrected for a $41\ \mu\mathrm{s}$  delay in the digital readout system, measured by triggering an LED with the PPS signal.  Barycentering was done with TEMPO2, using a custom plug-in that allows for the analysis of individual photons.  

The photons within a circular aperture (with a radius of five pixels) around the center of the pulsar's point spread function (PSF) were folded into pulse profiles, each with 250 phase bins, for GRP-accompanied optical pulses and their surrounding (40 pulses before and after) non-GRP-accompanied pulses.  The standard deviation of the flux of the surrounding non-GRP-accompanied pulses was used to estimate the error in the GRP-accompanied profile.  The sky level in the profiles was determined by the average of five phase bins in the off-pulse phase region and then subtracted.  Following \citet{ref:Shearer2003}, the optical flux enhancement was calculated as the percent increase in the number of photons in the three phase bins around the main peak.  

Spectra of the peak of the optical main pulse were made separately for GRP-accompanied and non-GRP-accompanied pulses.  We excluded the 20-30\% of pixels that did not have a reliable wavelength calibration in the range from 4000 to 11000 \AA ~(as determined by the wavelength calibration module of the ARCONS pipeline). The flat-field and spectral shape calibrations discussed in \citet{ref:Mazin2013} were not used for this work, so the spectra calculated are not corrected for the wavelength dependence in each MKID's quantum efficiency.  It is valid to compare spectra determined in this way, but the absolute spectral shape is not fully calibrated.  The sky spectrum was also determined as the spectrum of pixels in a half-annulus around the pulsar's PSF, where only half an annulus was used to avoid contamination from the Crab pulsar's stellar neighbor.  The resulting sky spectrum was subtracted from all pulsar spectra.

\section{Results}

\subsection{Optical-Radio Lag and Optical Enhancement}
\label{sec:optenh}

We observe the optical pulse to lead the radio pulse by $202\pm36\ \mu\mathrm{s}$.  This was determined as the time difference between the highest flux bins in the optical pulse profile (with 500 phase bins instead of our usual 250) and the radio pulse profile.  The uncertainty comes from the optical phase half-bin width, which is 0.001 in phase or $33.6\ \mu\mathrm{s}$, the radio phase half-bin width, which is 0.00024 in phase or $8.2\ \mu\mathrm{s}$, and the uncertainty in the radio timing from the dispersion measure (DM), which is ${\sim} 10\ \mu\mathrm{s}$.  The overall uncertainty was estimated as these three uncertainties added in quadrature.  Our measured optical-radio lag time is comparable to the recent measurements of \citet{ref:Slowikowska2009} and \citet{ref:Oosterbroek2008} at 231$\pm$68$\ \mu\mathrm{s}$ and 255$\pm$21$\ \mu\mathrm{s}$, respectively.

\begin{figure}[tbp]
\centering
\includegraphics*[width=0.475\textwidth]{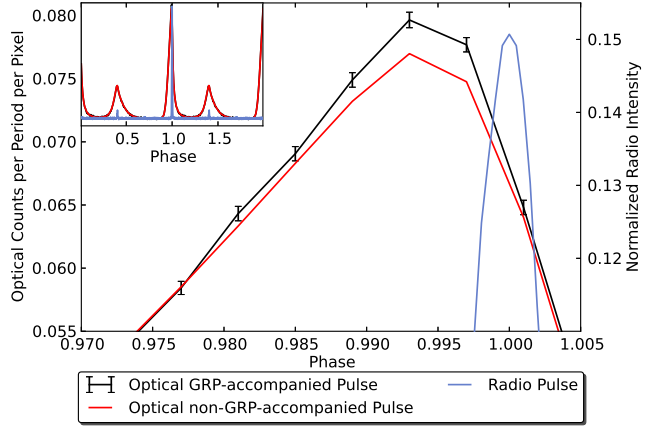}

\caption{
Average optical profile of the peak of the main pulse for 7205 pulses accompanied by a main pulse GRP (black) and for 80 pulses surrounding each of the 7205 accompanied pulses (excluding other GRP-accompanied pulses) (red).  The normalized radio pulse profile is also shown (blue). The inset shows two full pulse periods displaying the main pulses and the smaller interpulses in both optical (red) and radio (blue). 
}
\label{fig:enhancedProfile}
\end{figure}

\begin{figure}[tbp]
\centering
\includegraphics*[width=0.475\textwidth]{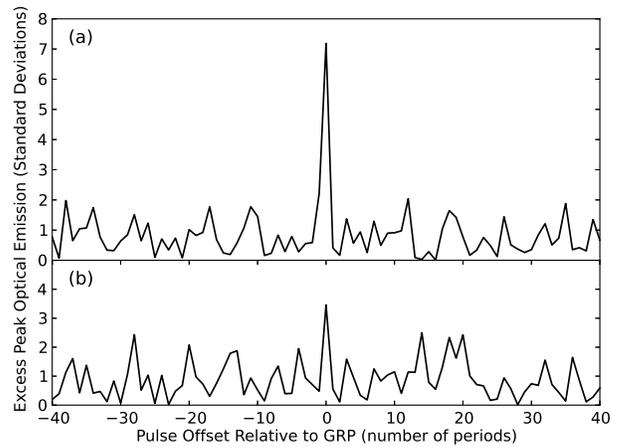}
\caption{
The number of standard deviations between the peak flux, for pulses with a fixed offset from a GRP, and the mean peak flux calculated for (a) main pulse GRPs and (b) interpulse GRPs.  Mean flux is an average over all pulses within 40 pulses of a GRP. The peak flux is defined as the sum of the three phase bins around the peak of the main pulse. 
}
\label{fig:significance}
\end{figure}

We found that when an optical pulse is accompanied by a main pulse GRP, the peak of the optical main pulse is enhanced, on average, by 3.2\%$\pm$0.5\% with a significance of 7.2 $\sigma$, as shown in Figure \ref{fig:enhancedProfile} and Figure \ref{fig:significance}.  Interpulse GRPs were observed to correlate with a moderately significant optical enhancement of the peak of the optical main pulse as well.  This enhancement was measured to be 2.8$\pm$0.8\% with a significance of 3.5 $\sigma$.  An enhancement of this kind was also observed by \citet{ref:Shearer2003} at 1.75 $\sigma$.  Our result is surprising given the expected number of false detections in the interpulse data. Future observations with higher signal to noise ratio may find higher enhancement.

\subsection{Radio Phase Correlations}

\begin{figure}[tbp]
\centering
\includegraphics*[width=0.475\textwidth]{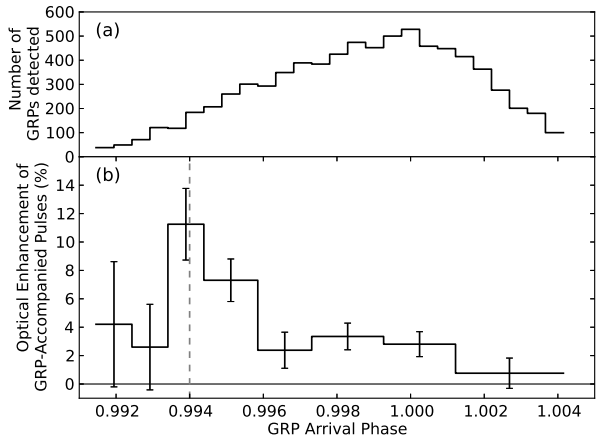}
\caption{
(a) Histogram of arrival phases for main pulse GRP detections.  The expected number of false positives is about 8 per arrival phase bin.  (b) Optical enhancement as a function of the GRP arrival phase.  For reference, the phase of the optical main pulse at phase 0.994 is shown.  
}
\label{fig:enhVsPhase}
\end{figure}

We found a correlation between optical enhancement and the phase of the accompanying main pulse GRP peak (Fig \ref{fig:enhVsPhase}).  The optical enhancement appears to be higher for certain arrival phases, with a maximum of 11.3$\pm$2.5\% for the phase bin with range 0.9934 to 0.9944, which is where the peak of the optical main pulse is.  To determine the significance of the observed enhancements diverging from a constant value, we performed a $\Delta \chi^2$ test.  In a similar application this test is sometimes used in the transit community to determine if a small eclipse is present in a light curve \citep{ref:Siverd2012_transit,ref:Fressin2011_transit}.  For the statistical test we use finer binning (bin widths are 0.0005 of a period).  We first fit a flat line model, resulting in a level of 3.22$\pm$0.46\%.  This fit had a $\chi^2$ of 40.5 with 25 degrees of freedom.   We then fit a Gaussian model with four parameters, amplitude, sigma, x-offset, and y-offset, which took on the values 8.3$\pm$2.4\%, 0.00075$\pm$0.00025, 0.99442$\pm$0.00025, and 2.54$\pm$0.54\%, respectively.  This resulted in a $\chi^2$ of 23.9 with 22 degrees of freedom.  Comparing the two fits gives a $\Delta \chi^2$ of 16.6 with 3 fewer degrees of freedom for the Gaussian model.  This difference corresponds to a p-value of 0.00084, indicating that the Gaussian model better represents the enhancement vs. phase data with a significance of 3.3 $\sigma$.

\subsection{Flux Correlations}
\label{sec:radioflux}

\begin{figure}[tbp]
\centering
\includegraphics*[width=0.475\textwidth]{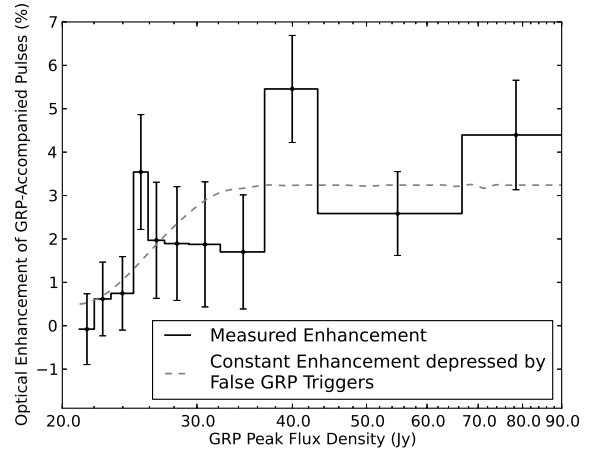}
\caption{
Enhancement of optical pulses accompanied by main pulse GRPs as a function of the peak flux density of the GRP (black).  The fraction of spurious radio peak detections increases as detected radio pulses become weaker.  The gray dashed line shows the predicted optical enhancement with the assumptions that optical pulses accompanied by false GRP detections have zero enhancement and that optical pulses accompanied by real GRPs have a 3.2\% enhancement that is constant with respect to GRP flux.}
\label{fig:enhVsRadioStrength}
\end{figure}

The results of probing for correlations between enhancement of optical pulses with an accompanying main pulse GRP and the flux of the GRP are shown in Figure \ref{fig:enhVsRadioStrength}.  In this plot, the optical enhancement appears to be lower for less energetic GRPs, but this is likely due to the treatment of spurious radio signals as GRPs.  Since spurious radio signals are not expected to have the enhancement associated with genuine GRPs and are more numerous at low flux densities, the calculated average enhancement is artificially lowered for lower flux bins.  The fraction of false counts was estimated as the ratio of the number of false GRP detections in an off-pulse phase range to the number of detections in the main pulse range.  Using this, Figure \ref{fig:enhVsRadioStrength} shows the expected enhancement curve if the actual enhancement is a constant of value 3.2\%.  Comparing the data curve with the predicted curve gives a $\chi^2$ of 9.5 with 10 degrees of freedom and corresponding p-value of 0.48, equivalent to 0.70 $\sigma$ significance in the difference between the data curve and predicted curve. Consequently, we find no significant change in the optical enhancement as a function of the flux density of the GRP.

\subsection{Spectral Dependence}

\begin{figure}[tbp]
\centering
\includegraphics*[width=0.475\textwidth]{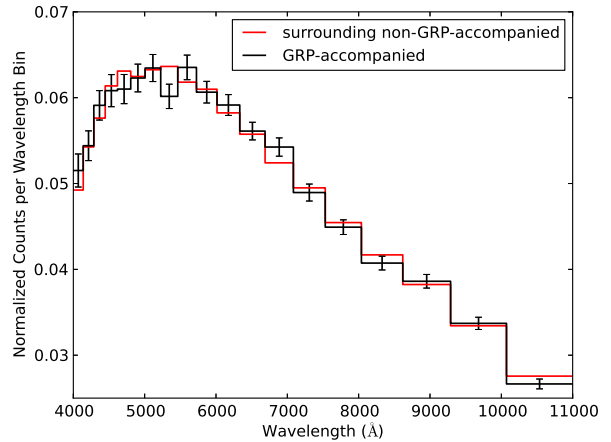}
\caption{
Spectrum of photons arriving during the peak of the optical main pulse (3 phase bins with highest counts) for GRP-accompanied pulses (black) and surrounding pulses (red) normalized to have the same integrated flux.  The wavelength resolution has been oversampled.  There do not appear to be significant spectral differences between enhanced GRP-accompanied pulses and non-GRP-accompanied pulses.
}
\label{fig:spectra}
\end{figure}

We found no statistically significant differences between the spectrum of the peak of the main pulse for enhanced optical pulses accompanied by a main pulse GRP and the spectrum for non-GRP-accompanied pulses (See Figure \ref{fig:spectra}). We performed a two-sample Kolmogorov-Smirnov (K-S) test to compare the spectral distribution of photons in GRP-accompanied and non-GRP-accompanied pulses with the null hypothesis that the photon wavelengths are drawn from the same spectral distribution in both cases.  The result was a D-metric of 0.0030 with a corresponding p-value of 0.41.  So, any difference between the spectra has a significance of only 0.83 $\sigma$.  

\section{Discussion}

The apparent match between the spectrum of ordinary pulses and GRP-enhanced optical pulses gives us a clue about the nature of the relationship between the radio and optical emission of the pulsar.  In order to produce the same spectrum, the mechanism that generates the extra optical emission during an enhanced pulse is likely the same mechanism that generates ordinary optical emission.

Interestingly, around the arrival phase for which GRPs were correlated with higher enhancements of the peak of the optical main pulse, \citet{ref:Slowikowska2009} found that the slope of the average optical polarization angle changes rapidly.

%Interestingly, Lommen et al. \cite{ref:Lommen2007} found higher flux density in the X-ray profile of the Vela pulsar, in the X-ray peak immediately following the radio pulse, when the radio pulse arrives early.

The fact that the degree of optical enhancement depends on the phase of the corresponding GRP for the Crab pulsar strengthens the link between optical and radio emission.  \citet{ref:Shklovsky1970} proposed that radio photons catalyze optical emission from pulsars. He noted that the large electric fields generated by the rotation of the pulsar's magnetic field easily accelerate electrons and positrons to high energies along field lines, but cannot easily contribute momentum perpendicular to the enormous B-field of the star. However, in the frame of a relativistic particle, a radio photon may be blueshifted to the frequency of the cyclotron resonance, and convert some of the particle's momentum along a magnetic field line to the transverse direction. The particle then emits synchrotron radiation at optical to X-ray energies. The spectral dissection of \citet{ref:Harding2008} ascribes spectral emission of the Crab pulsar at these energies to synchrotron emission, and notes the role of catalysis by radio photons. 

Shklovsky's picture would suggest that the stimulating radio pulse should arrive with the resulting optical emission.  We observe this sequence for the most enhanced optical pulses, but often the radio pulse arrives later than the enhanced optical pulse, suggesting that a less direct mechanism is at work. The radio pulse may suffer significant delay due to magnetospheric refraction \citep{ref:Barnard1986,ref:Lyutikov2000,ref:Jessner2010,ref:Beskin2012}.  Alternatively, if optical and radio emission were beamed in slightly different directions, they would appear at different pulse phase. However, this would require the duration of GRPs to be at least as long as the optical-radio lag.  This duration would be difficult to reconcile with the discovery by \citet{ref:Hankins2003} that GRPs are made of nanosecond bursts separated by microseconds.

\section*{Conclusion}

Utilizing the unique abilities of MKIDs, we have shown that the enhancement of optical pulses accompanied by a main pulse GRP has a significant dependence on the arrival phase of the GRP.  Our data also shows some correlation between enhancement of optical main pulses and interpulse GRPs.  In addition, the spectra of enhanced optical pulses and normal pulses do not differ significantly, and no significant relationship is detected between optical enhancement and GRP flux.  Future observations with ARCONS should improve the signal-to-noise ratio in probing the arrival phase to flux relationship in order to better resolve the apparent peak.  ARCONS will also be used to observe other, much fainter rotation-powered pulsars and to search for optical pulsations in millisecond pulsars.

\acknowledgments
The MKID detectors used in this work were developed under NASA grant NNX11AD55G.  The MKID digital readout was partially developed under NASA grant NNX10AF58G.  SRM was supported by a NASA Office of the Chief Technologist's Space Technology Research Fellowship, NASA grant NNX11AN29H.  This work was partially supported by the Keck Institute for Space Studies.  CG, MJ, and GVSJ thank the U.S. National Science Foundation for financial support for this work (AST-1008865).  Fermilab is operated by Fermi Research Alliance, LLC under Contract No. De-AC02-07CH11359 with the United States Department of Energy.  

The authors would like to thank Shri Kulkarni, Director of the Caltech Optical Observatories, and Tom Prince for facilitating this project, and would like to thank Paul Demorest for his assistance with the GBT observations.  Also, the authors thank Jason Eastman for fruitful correspondence.

%\appendix

{\it Facilities:} \facility{GBT (GUPPI)}, \facility{Hale} %my version needs a line break to build properly

\end{document}